\documentclass[twocolumn,showpacs,preprintnumbers,prb,amsmath,amssymb]{revtex4}
\usepackage{graphicx}% Include figure files
\usepackage{dcolumn}% Align table columns on decimal point
\usepackage{bm}% bold math

\begin{document}

\title{Avoiding power broadening in optically detected magnetic resonance\\
 of single NV defects for enhanced DC-magnetic field sensitivity}

\date{\today}

\author{A. Dr\'eau$^{1}$}
\author{M. Lesik$^{1}$}
\author{L. Rondin$^{1}$}
\author{P. Spinicelli$^{1}$}
\author{O. Arcizet$^{2}$}
\author{J.-F. Roch$^{1}$}
\author{V. Jacques$^{1}$}
\affiliation{$^{1}$Laboratoire de Photonique Quantique et Mol\'eculaire, Ecole Normale Sup\'erieure de Cachan and CNRS, UMR 8537, 94235 Cachan, France}
\affiliation{$^{2}$Institut N\'eel, Universit\'e Joseph Fourier and CNRS, UPR 2940, 38042 Grenoble, France}
\date{\today}

\begin{abstract}
We report a systematic study of the magnetic field sensitivity of a magnetic sensor consisting of a single Nitrogen-Vacancy (NV) defect in diamond, by using continuous optically detected electron spin resonance (ESR) spectroscopy. We first investigate the behavior of the ESR contrast and linewidth as a function of the microwave and optical pumping power. The experimental results are in good agreement with a simplified model of the NV defect spin dynamics, leading to an optimized sensitivity around $2 \mu$T$/\sqrt{\rm Hz}$ for a single NV defect in a high purity CVD-grown diamond crystal. We then demonstrate an enhancement of the magnetic sensitivity by one order of magnitude by using a simple pulsed-ESR scheme. This technique is based on repetitive excitation of the NV defect with a resonant microwave $\pi$-pulse followed by an optimized read-out laser pulse, allowing to fully eliminate power broadening of the ESR linewidth. The achieved sensitivity is similar to the one obtained by using Ramsey-type sequences, which is the optimal magnetic field sensitivity for the detection of a DC magnetic field.
\end{abstract}


\maketitle

\section{Introduction}
Among many studied optically-active defects in diamond~\cite{Zaitsev}, the negatively-charged Nitrogen-Vacancy (NV) color center has attracted a lot of interest during the past years owing to its unprecedented optical and spin properties at room temperature~\cite{Jelezko_Revue2006}. The NV defect electron spin can be initialized, coherently manipulated with long coherence time~\cite{Jelezko_PRL2004,Gopi_NatMat2009} and read-out by pure optical means through its perfectly photostable spin-dependent photoluminescence (PL). Such properties are the cornerstone of a wide range of emerging technologies, from imaging in life science~\cite{Hollenberg_NatNano2011}, to quantum information processing~\cite{Dutt_Science2007,Neumann_NatPhys2010,Neumann_Science2010,Kubo_PRL2010,Togan_Nature2010,Buckley_Science2010} and high resolution sensing of magnetic~\cite{Gopi_Nature2008,Maze_Nature2008,Steinert_RSI2010,Pham_NJP2011,Maertz_APL2010,Harneit_PRL2011} and electric fields~\cite{Dolde_NatPhys2011}.\\
\indent For magnetometry applications, the principle of the measurement is similar to the one used in optical magnetometers based on the precession of spin-polarized atomic gases~\cite{Cohen_PLA1969,Romalis_NatPhys2007}. The applied magnetic field is evaluated through the detection of Zeeman shifts of the NV defect spin sublevels. The associated magnetic field sensitivity has been thoroughly analyzed both theoretically~\cite{Taylor_NatPhys2008,Meriles_JCP2010} and experimentally, by using either Ramsey-type pulse sequences for the detection of a DC magnetic field~\cite{Gopi_NatMat2009}, or dynamical decoupling sequences for AC magnetic field sensing~\cite{Maze_Nature2008,Laraoui_APL2010,Hanson_PRL2011}. \\
\indent Although ultrahigh sensitivity can be achieved using 
multi-pulse sensing sequences, the simplest way to measure an external DC magnetic field with a single NV defect remains the direct evaluation of the Zeeman splitting in an optically detected electron spin resonance (ESR) spectrum. In this paper, we focus on that simple case. By using a simplified model of the NV defect spin dynamics, we first investigate how the magnetic field sensitivity evolves with the microwave power and the optical pumping power, respectively used for spin rotations and spin polarization in continuous optically detected ESR spectroscopy. We then demonstrate a pulsed-ESR method which allows to fully eliminate power broadening of the ESR linewidth~\cite{Vitanov_OptCom2001}. This technique uses repetitive excitation of the NV defect with a resonant microwave $\pi$-pulse followed by an optimized read-out laser pulse, and leads to an enhancement of the magnetic sensitivity by one order of magnitude. The achieved sensitivity is similar to the one obtained with Ramsey-type sequences, which corresponds to the optimal magnetic field sensitivity for the detection of a DC magnetic field.

\section{Magnetic field sensitivity using continuous ESR spectroscopy}
\subsection{The NV defect in diamond as a magnetic sensor}

\begin{figure}[t]
 \includegraphics[width = 8.7cm]{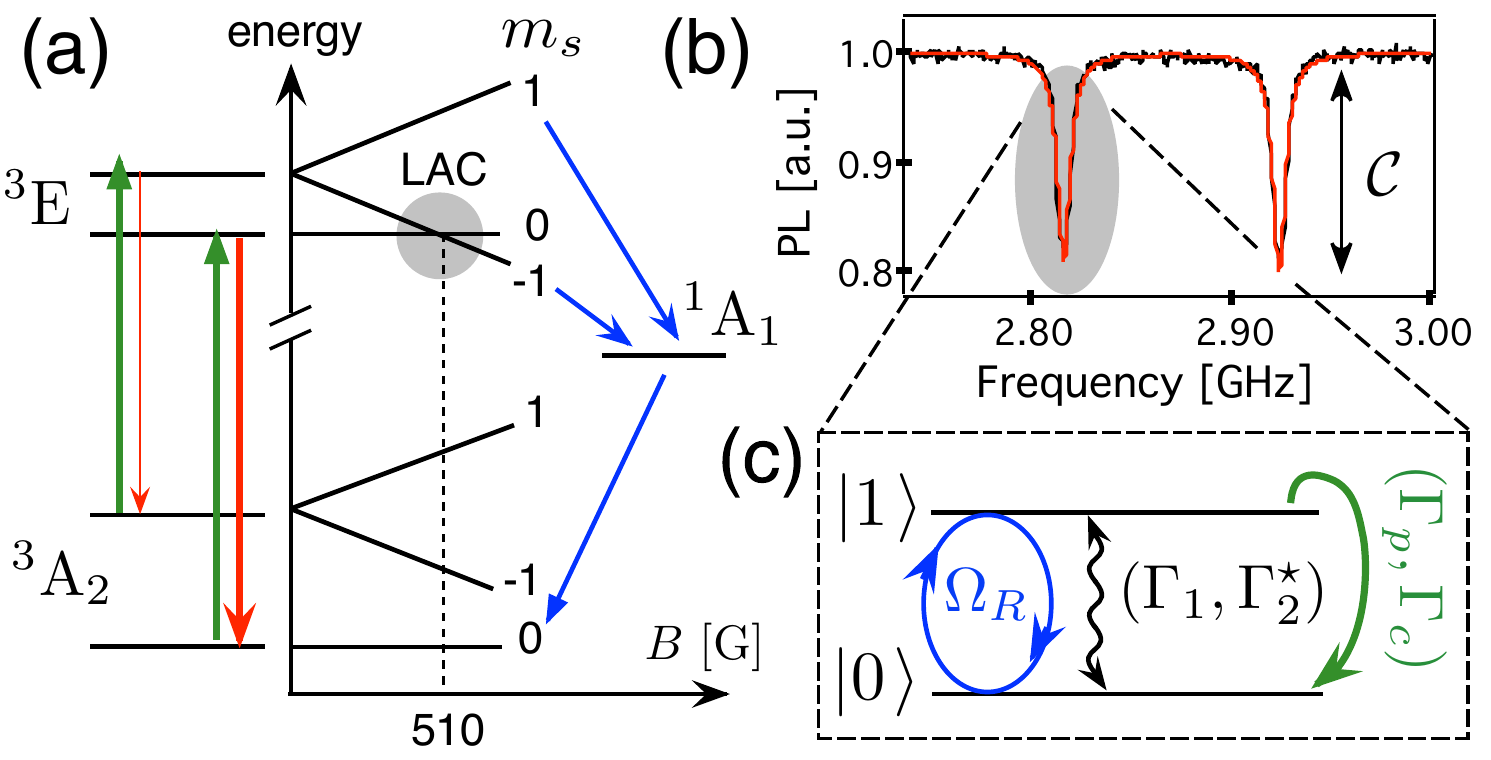}
\caption{(color online). (a)-Energy-level diagram of the NV defect as a function of the amplitude of a static magnetic field $B$ applied along the NV defect axis. The ground state $^{3}{\rm A}_{2}$ and the excited state $^{3}$E are electron spin triplets. Optical transitions $^{3}{\rm A}_{2}\rightarrow ^{3}$E are spin-conserving while non-radiative ISC to an intermediate singlet state $^{1}{\rm A}_{1}$ is strongly spin dependent (blue arrows). The grey circle indicates the excited-state level anti-crossing (LAC), achieved for a magnetic field amplitude $B\approx 510$ G applied along the NV axis. (b)-Typical ESR spectrum of a single NV defect recorded with a static magnetic field $B\approx 20$~G applied along the NV axis. Solid line is data fitting using Lorentzian functions.(c)-The NV defect spin dynamics is studied by considering a two-level system in interaction with a resonant microwave field. Notations are defined in the main text.}
 \label{Fig1}
 \end{figure}

\indent  The negatively-charged NV defect in diamond consists of a substitutional nitrogen atom (N) associated with a vacancy (V) in an adjacent lattice site of the diamond matrix. This defect exhibits an efficient and perfectly photostable red photoluminescence (PL), which enables optical detection of individual NV defects by confocal microscopy at room temperature~\cite{Gruber_Science1997}. The NV defect ground state is a spin triplet with $^{3}$A$_{2}$ symmetry, whose degeneracy is lifted by spin-spin interaction into a singlet state of spin projection $m_{s}=0$ and a doublet $m_{s}=\pm 1$, separated by $2.87$~GHz in the absence of magnetic field (Fig.~\ref{Fig1}(a)). Spin-conserving optical transitions $^{3}{\rm A}_{2}\rightarrow ^{3}$E combined with spin-selective intersystem-crossing (ISC) towards an intermediate singlet state $^{1}{\rm A}_{1}$ provide a high degree of electron spin polarization in the $m_{s}=0$ sublevel through optical pumping~\cite{Manson_PRB2006,Maze_NJP2011}. Furthermore, the NV defect PL intensity is significantly higher (up to $\approx 20\%$) when the $m_{s}=0$ state is populated. This spin-dependent PL response enables the detection of electron spin resonance (ESR) on a single defect by pure optical means~\cite{Gruber_Science1997}. \\
 \indent We investigate native NV defects in a ultra-pure synthetic type IIa diamond crystal prepared using a microwave assisted chemical vapor deposition (CVD) growth (Element6). Individual NV defects are optically addressed at room temperature using a confocal optical microscope. A laser operating at $532$ nm wavelength is focused onto the sample through a high numerical aperture oil-immersion microscope objective (Olympus, $\times 60$, NA=1.35). The NV defect PL is collected by the same objective, focused onto a 50-$\mu$m-diameter pinhole and finally directed to a photon counting detection system. In addition, a microwave field is applied through a copper microwire directly spanned on the diamond surface and a weak static magnetic field is applied along the NV axis in order to lift the degeneracy of the $m_{s}=\pm 1$ spin sublevels. ESR spectroscopy of single NV defects is performed by sweeping the frequency of the microwave field while monitoring the PL intensity. When the microwave frequency is resonant with the transition between $m_{s}=0$ and one of the $m_{s}=\pm1$ states, spin rotation is evidenced as a dip of the PL signal (Fig.~\ref{Fig1}(b)). Using such a spectrum, the NV defect can be used as a nanoscale magnetic sensor by measuring Zeeman shifts of the ESR line induced by a remote magnetic field~\cite{Gopi_Nature2008,Maze_Nature2008}.
 
 \subsection{Magnetic field sensitivity}
The intensity $\mathcal I$ of optically detected ESR spectra as a function of the microwave frequency $\nu_{m}$ can be written as 
\begin{equation}
\mathcal I(\nu_{m})={\mathcal R}\left[1-\mathcal C \mathcal{F}\left(\frac{\nu_{m}-\nu_0}{\Delta\nu}\right)\right] \ ,
\end{equation}
where $\mathcal R$ is the rate of detected photons, $\mathcal C$ the ESR contrast associated to the dip of the PL intensity, $\mathcal{F}$ the ESR lineshape and $\Delta\nu$ the associated linewidth (FWHM). Any magnetic field fluctuation $\delta B$ induces a shift of the central frequency $\nu_{0}$ through the Zeeman effect. The photon shotnoise of a measurement with $\Delta t$ duration has a standard deviation $\sqrt{\mathcal I \Delta t}$. For low ESR contrast, the shotnoise-limited magnetic field sensitivity $\eta_{B}$ of this measurement is linked to the minimum detectable magnetic field $\delta B_{\rm min}$ through the relation~\cite{Taylor_NatPhys2008,Harneit_PRL2011}
\begin{equation}
\eta_B(\mathrm {T}/\sqrt{\mathrm{Hz}})=\delta B_{\rm min}\sqrt{\Delta t}\approx \frac{h}{g\mu_{B}}\times\frac{\sqrt{\mathcal R}}{\displaystyle\max\lvert\frac{\partial \mathcal I}{\partial \nu_{0}}\rvert} \ ,
\end{equation}
which reads as
\begin{equation}
\eta_B\approx \mathcal P_\mathcal{F} \times \frac{h}{g\mu_B}\times\frac{\Delta\nu}{\mathcal C\sqrt{\mathcal R}} \ .
\label{sensib}
\end{equation}
In this expression, $\mathcal P_\mathcal{F}$ is a numerical parameter related to the specific profile $\mathcal{F}$ of the spin resonance. For a Gaussian profile, $\mathcal P_G=\sqrt{e/8\ln 2}\approx0.70$ whereas a Lorentzian profile leads to $\mathcal P_L=4/3\sqrt3\approx0.77 $.

\indent The ESR linewidth $\Delta\nu$ is fundamentally limited by the inhomogeneous dephasing rate $\Gamma_{2}^{*}$ of the NV defect electron spin, which is determined by magnetic dipolar interactions with a bath of spin impurities inside the diamond matrix. In high purity CVD-grown diamond crystal considered in this study, these impurities are essentially the nuclear spins associated with carbon isotope $^{13}$C ($I=1/2$, natural abundance $1 \%$). The effect of the nuclear-spin bath can be interpreted as a randomly fluctuating magnetic field applied to the central single spin. In the limit of a large number of bath spins, the distribution of this effective magnetic field is determined by the central limit theorem to be Gaussian~\cite{Dobrovitski_PRB2008}. However, the ESR linewidth is also affected by power broadening, both from the continuous laser light used for spin polarization and from the resonant microwave field used for spin rotation. This results in a power broadened Lorentzian profile of the ESR linewidth. In order to sharpen the ESR linewidth, corresponding to an enhanced magnetic field sensitivity, the microwave power and the laser intensity need to be decreased. However, the drawback of this simple method is a significant reduction of the ESR contrast and of the rate $\mathcal{R}$ of detected photons, which impair the magnetic field sensitivity as analysed in the next section. 

\subsection{Simplified model of NV defect spin dynamics}

We first develop a toy model of the NV defect spin dynamics in order to infer the behavior of the contrast and the linewidth of the ESR as a function of the microwave and the optical pumping power. For that purpose, we consider the NV defect as a simple {\it closed} two-level system, denoted $\left|0\right.\rangle$ and $\left|1\right.\rangle$ and respectively corresponding to the ground states with spin projection $m_{s}=0$ and $m_{s}=-1$ (Fig.~\ref{Fig1}(c)). The Hamiltonian $\mathcal{H}$ describing the interaction of the system with a quasi-resonant magnetic field oscillating at the microwave frequency $\omega_{m}$ reads as 
\begin{equation}
\hat{\mathcal{H}}=\hbar\omega_{0}\left|1 \rangle \langle 1 \right | + \hbar\Omega_{R} \cos(\omega_{m} t) \left( \left|0 \rangle \langle 1 \right | + \left|1 \rangle \langle 0 \right | \right) \ ,
\label{Hamilto}
\end{equation}
where $\omega_{0}$ is the Bohr frequency of the spin transition and $\Omega_{R}$ the Rabi frequency of the magnetic dipole interaction. Using the formalism of the density operator $\hat{\sigma}$, the evolution of the system is then described by the Liouville equation
\begin{equation}
\frac{d\hat{\sigma}}{dt}=\frac{1}{i\hbar}\left[\hat{\mathcal{H}},\hat{\sigma}\right]+\Big\{\frac{d\hat{\sigma}}{dt}\Big\}_{\rm relax} \ ,
\label{Liouville}
\end{equation}
where the last term describes the relaxation of the system through its interaction with the environment. Following nuclear magnetic resonance terminology, intrinsic relaxation of the populations $\sigma_{ii}$ occurs through spin-lattice relaxation process with a rate $\Gamma_{1}$, while coherences $\sigma_{ij}$ decay with an inhomogeneous dephasing rate $\Gamma_{2}^{*}$. Typical values for a single NV defect in a high purity CVD-grown diamond are~\cite{Gopi_NatMat2009,Nori_PRB2009} $\Gamma_{1}\approx10^{3}$~s$^{-1}$ and $\Gamma_{2}^{*}\approx 2\times 10^{5}$~s$^{-1}$.\\
\indent Within this simplified framework, we do not consider populations neither in the excited state nor in the metastable state. The effect of optical pumping is thus phenomenologically introduced through an induced relaxation process, both for populations and coherences. The effect of the metastable state responsible for spin polarization is described by a relaxation process of the population $\sigma_{11}$ with a polarization rate $\Gamma_{p}$. Since spin-selective ISC to the metastable state is induced by spin-conserving optical transitions (Fig.~\ref{Fig1}(a)), the polarization rate $\Gamma_{p}$ is related to the rate of optical cycles, which follows a standard saturation behavior with the optical pumping power $\mathcal{P}_{\rm opt}$. Denoting $\mathcal{P_{\rm sat}}$ the saturation power of the transition,  the optically-induced polarization rate $\Gamma_{p}$ can be given by
\begin{equation}
\Gamma_{p}=\Gamma_{p}^{\infty}\times \frac{s}{1+s} \ ,
\label{gamP}
\end{equation}
where $s=\mathcal{P}_{\rm opt}/\mathcal{P_{\rm sat}}$ is the saturation parameter of the radiative transition and $\Gamma_{p}^{\infty}$ the polarization rate at saturation. This quantity is fixed by the lifetime of the metastable state, which is on the order of $200$ ns at room temperature~\cite{Hanson_NJP2011}, leading to $\Gamma_{p}^{\infty}\approx 5\times10^{6}$~s$^{-1}$.   \\
\indent Optical pumping also leads to relaxation of the electron spin coherences $\sigma_{ij}$. Since only a few scattered photons are enough to destroy the phase information, the relaxation rate of coherences induced by optical pumping can be written as
\begin{equation}
\Gamma_{c}=\Gamma_{c}^{\infty}\times \frac{s}{1+s} \ ,
\label{gam2}
\end{equation}
where $\Gamma_{c}^{\infty}$ is the rate of optical cycles at saturation. This quantity is set by the excited-state radiative lifetime, which is on the order of $13$ ns~\cite{Manson_PRB2006}, leading to $\Gamma_{c}^{\infty}\approx 8\times 10^{7}$~s$^{-1}$. \\
\indent By including the intrinsic ($\Gamma_{1}$,$\Gamma_{2}^{\star}$) and the optically-induced ($\Gamma_{p}$,$\Gamma_{c})$ relaxation processes in the Liouville equation~(Eq.~(\ref{Liouville})), the steady state solutions of the system $\sigma_{ii}^{st}$ can be easily computed (see Appendix). The NV defect PL rate can then be written as 
\begin{equation}
\mathcal{R}(\Omega_{R},\omega_{m},s)=\left[\alpha\sigma_{00}^{st}+\beta\sigma_{11}^{st}\right]\times \displaystyle\frac{s}{1+s} \ ,
\label{PL}
\end{equation}
where the parameters $\alpha$ and $\beta$ are phenomenologically introduced in order to account for the difference in PL intensity between the $m_{s}=0$ and $m_{s}=1$ spin sublevels ($\alpha>\beta$). Using above notations, the ESR contrast can then be evaluated as (see Fig.~\ref{Fig1}(b))
\begin{equation}
\mathcal{C}=\frac{\mathcal{R}(0,0,s)-\mathcal{R}(\Omega_{R},\omega_{0},s)}{\mathcal{R}(0,0,s)} \ ,
\end{equation}
where $\mathcal{R}(0,0,s)$ (resp. $\mathcal{R}(\Omega_{R},\omega_{0},s)$) denotes the NV defect PL rate without applying the microwave field (resp. with a resonant microwave field). \\
\indent The general derivation of the contrast is given in Appendix. In the following, we assume that optical pumping is such that $s>10^{-2}$, which is usual for experiments aiming at the optical detection of single NV defects. Intrinsic relaxation processes can then be neglected, {\it i.e.} $\Gamma_{p}\gg \Gamma_{1}$ and $\Gamma_{c}\gg \Gamma_{2}^{\star}$, and the ESR contrast simply reads as
\begin{equation}
\mathcal{C}=\Theta \times \frac{\Omega_{R}^{2}}{\Omega_{R}^{2}+\Gamma_{p}^{\infty}\Gamma_{c}^{\infty}\displaystyle \left(\frac{s}{1+s}\right)^{2}} \, 
\label{Cfinal}
\end{equation}
where $\Theta=(\alpha-\beta)/2\alpha$ appears as an overall normalization factor (see Appendix). The ESR contrast evolves in opposite ways with respect to the optical pumping power and to the amplitude of the microwave field. Indeed, for a fixed value of the Rabi frequency $\Omega_{R}$, the ESR contrast increases while decreasing the optical pumping power. On the other hand, for a fixed saturation parameter $s$, the ESR contrast drops while decreasing the Rabi frequency, {\it i.e.} the microwave power.\\
\indent The other important parameter taking part in the magnetic field sensitivity is the ESR linewidth (see eq.~(\ref{sensib})). Within Bloch equation formalism described above and assuming $s>10^{-2}$, the ESR exhibits a Lorentzian profile with a power-broadened linewidth $\Delta\nu$ (FWHM) given by 
\begin{equation}
\label{Linewidth}
\Delta\nu=\frac{\Gamma_{c}^{\infty}}{2\pi}\times \sqrt{\displaystyle (\frac{s}{1+s})^{2}+\frac{\Omega_{R}^{2}}{\Gamma_{p}^{\infty}\Gamma_{c}^{\infty}}} . 
\end{equation}

\subsection{Experimental results} 

\begin{figure}[t]
\includegraphics[width = 8.8cm]{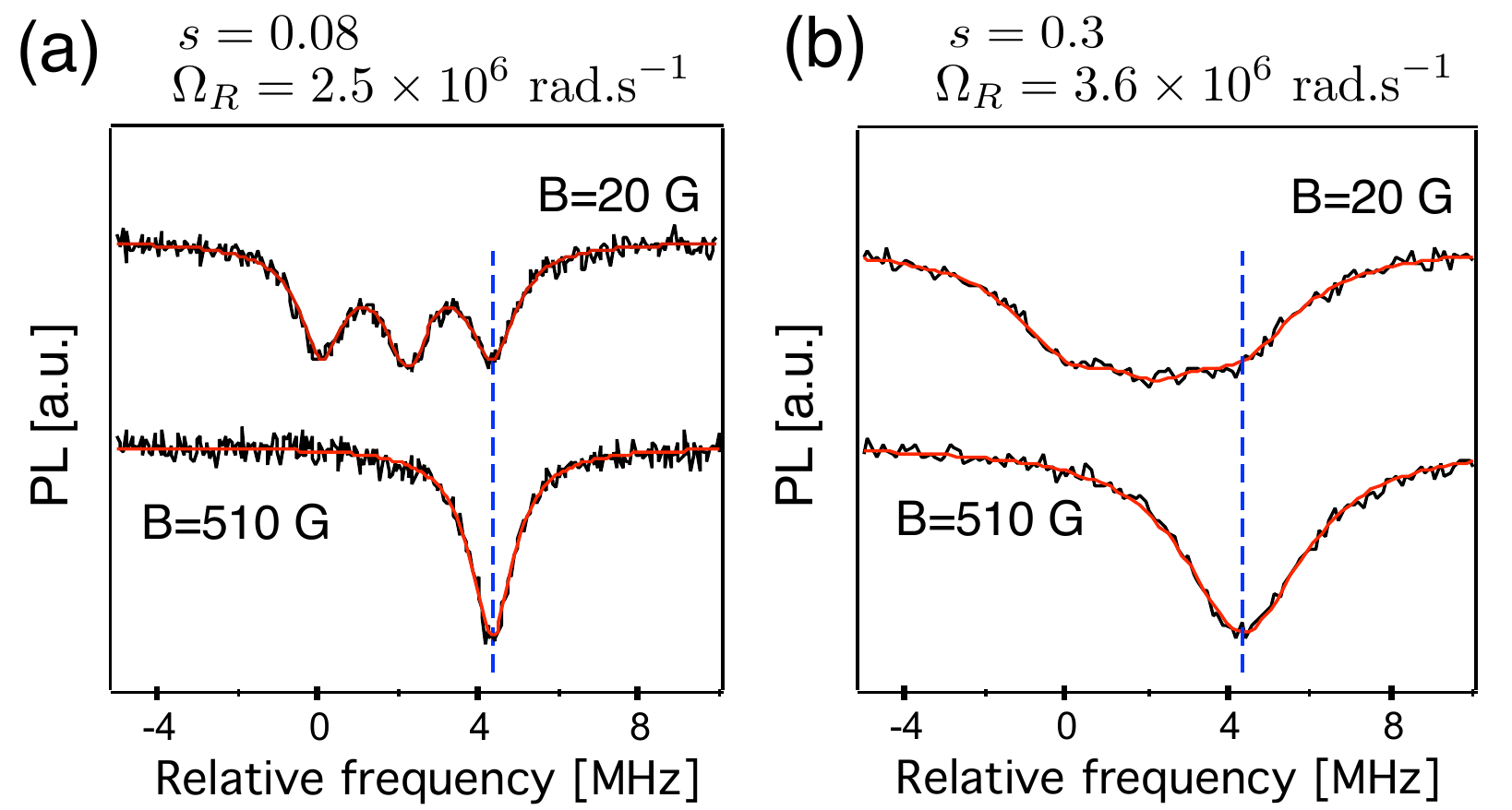}
\caption{(color online). (a) and (b)-ESR spectra recorded for a magnetic field amplitude $B=20$~G (upper traces) and at the excited-state LAC for $B\approx 510$~G applied along the NV defect axis (bottom traces). Since the $^{14}$N nuclear spin is fully polarized at the excited-state LAC, unambiguous measurements of the contrast and linewidth of the ESR can be obtained. Solid lines are data fitting using Lorentzian functions. Data displayed in (a) (resp.~(b)) are recorded for $s=0.08$ (resp. $s=0.3$) and $\Omega_{R}=2.5\times 10^{6}$~rad.s$^{-1}$ (resp. $\Omega_{R}=3.6\times 10^{6}$~rad.s$^{-1}$). }
 \label{ESLAC}
 \end{figure}

We now check experimentally how the contrast and the linewidth of the ESR evolve with the microwave and the optical pumping power.\\
\indent If the nitrogen atom of the NV defect is a $^{14}$N isotope ($99.6 \%$ abundance), corresponding to a nuclear spin $I=1$, each electron spin state is split into three sublevels by hyperfine interaction. ESR spectra thus exhibit three hyperfine lines, splitted by $2.16$ MHz~\cite{Felton_PRB2009} and corresponding to the three nuclear spin projections (Fig.~\ref{ESLAC}(a), top trace). When the ESR linewidth is larger than the hyperfine splitting, the three lines add up leading to uncertainties in the measurement of the contrast and the linewidth of each single line (Fig.~\ref{ESLAC}(b), top trace). To circumvent this problem, all the measurements were performed at the excited state level anti-crossing (LAC), while applying a static magnetic field near $510$ G along the NV axis~\cite{Fuchs_PRL2009,Neumann_NJP2009} (Fig.~\ref{Fig1}(a)). In this configuration, electron-nuclear-spin flip-flops mediated by hyperfine interaction in the excited-state lead to an efficient polarization of the $^{14}$N nuclear spin~\cite{Jacques_PRL2009,Childress_PRA2009,Steiner_PRB2010}. As a result, a single ESR line is observed at the excited-state LAC, leading to unambiguous measurement of the ESR contrast and linewidth (Fig.~\ref{ESLAC}(a) and (b), bottoms traces).
 
  \begin{figure}[t]
 \includegraphics[width = 9cm]{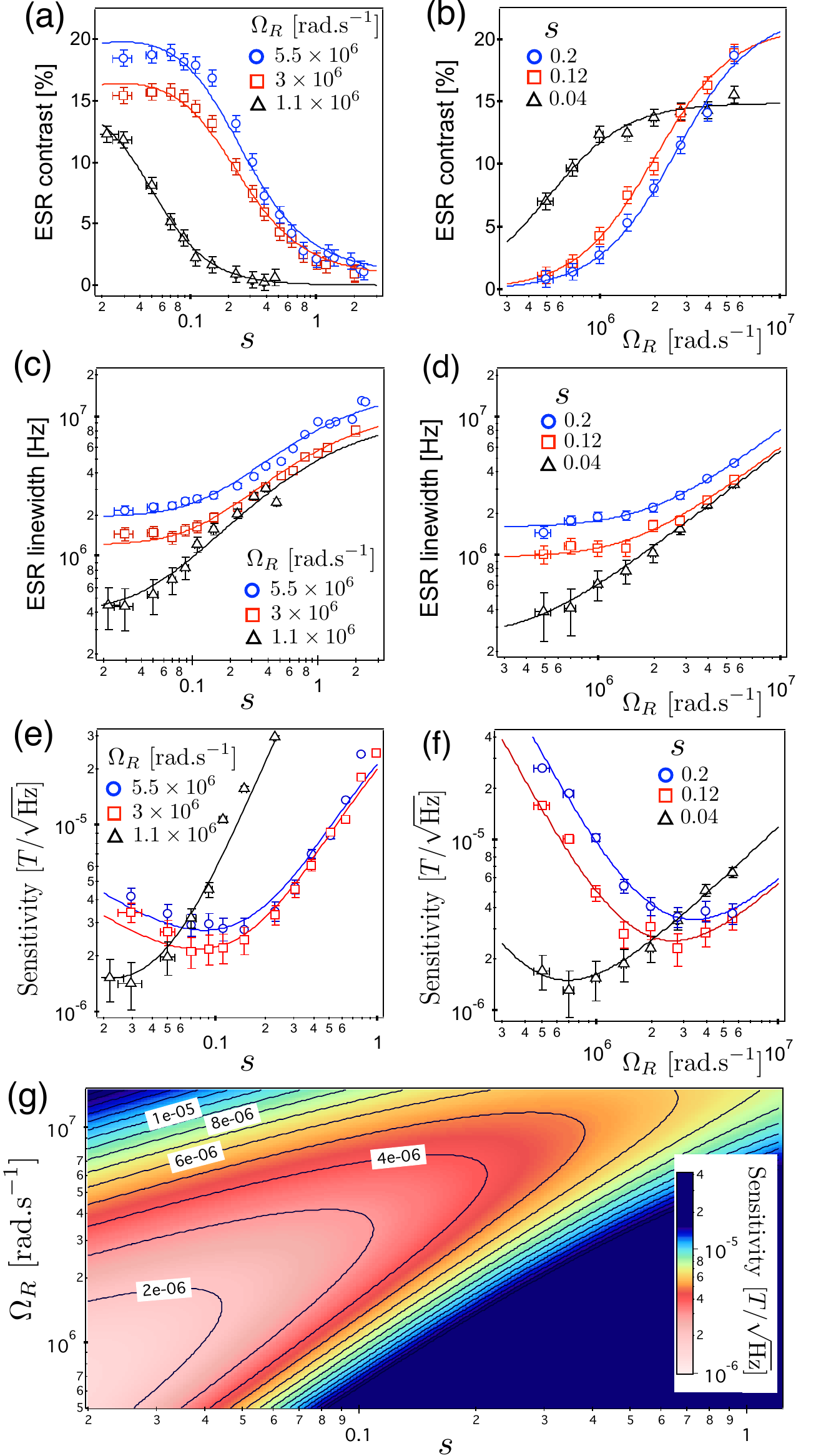}
\caption{(color online). (a) to (d)-ESR contrast and linewidth as a function of the Rabi frequency $\Omega_{R}$ and the saturation parameter $s$ plotted in log scale. The solid lines are data fitting using eqs.~(\ref{Cfinal}) and~(\ref{Linewidth}) as discussed in the main text. (e),(f)- Corresponding magnetic field sensitivities plotted in log-log scale using eq.~(\ref{sensib}). (g)-Two dimensional plot of the magnetic field sensitivity obtained by using eqs.~(\ref{sensib}),~(\ref{Cfinal}),~(\ref{Linewidth}) and the results of data fitting: $\Gamma_{p}^{\infty}=5\times 10^{6}$~s$^{-1}$, $\Gamma_{c}^{\infty}=8\times 10^{7}$~s$^{-1}$, $\Theta=0.2$ and $\mathcal{R}^{\infty}=250 \times 10^{3}$ counts.s$^{-1}$. The solid lines correspond to iso-magnetic field sensitivities.}
 \label{FigMod1}
 \end{figure}
\indent These parameters were measured while changing the saturation parameter $s$ and the Rabi frequency $\Omega_{R}$ (Fig.~\ref{FigMod1}(a) to (d)). The latter was independently measured by recording electron spin Rabi oscillations using the standard pulse sequence described in Ref.~\cite{Jelezko_PRL2004}. As anticipated, the ESR contrast increases while decreasing the optical pumping power, and it lowers while decreasing the strength of the microwave field. Measurements of the contrast are well fitted using eq.~(\ref{Cfinal}), with $\Theta$ and $\Gamma_{p}^{\infty}\Gamma_{c}^{\infty}$ as fitting parameters (Fig.~\ref{FigMod1}(a) and (b)). On the other hand, the ESR linewidth decreases both with the microwave and the optical pumping power, as expected from usual power broadening. Once again the measurements are in reasonable agreement with the model as illustrated by data fitting using eq.~(\ref{Linewidth}) with $\Gamma_{p}^{\infty}$ and $\Gamma_{c}^{\infty}$ as fitting parameters (Fig.~\ref{FigMod1}(c) and (d)).  Although the simplified model of the NV defect spin dynamics developed in this study does not allow to extract precise values of the photophysical parameters, the results of the fits give on average $\Gamma_{p}^{\infty}=6\pm 2\times 10^{6}$~s$^{-1}$ and $\Gamma_{c}^{\infty}=8\pm 2 \times 10^{7}$~s$^{-1}$, which are of good order-of-magnitude as discussed in the previous section. The two-level toy model  is thus sufficient to explain the behavior of the ESR contrast and linewidth as a function of the microwave and the optical pumping power.\\
\indent According to eq.~(\ref{sensib}), the last parameter required to infer the magnetic field sensitivity $\eta_{B}$ is the rate of detected photons $\mathcal{R}$. This parameter follows a saturation behavior
\begin{equation}
\label{Sat}
\mathcal{R}=\mathcal{R}^{\infty}\frac{s}{1+s} \ ,
\end{equation}
where $\mathcal{R}^{\infty}$ is the rate of detected photons at saturation. In our experimental setup, we measured $\mathcal{R}^{\infty}=250\times 10^{3}$ counts.s$^{-1}$ for single NV defects with a typical saturation power $\mathcal{P_{\rm sat}}\approx 250 \ \mu$W (data not shown).\\
\indent From this set of measurements, the magnetic field sensitivity $\eta_{B}$ was estimated using eq.~(\ref{sensib}) (Fig.~\ref{FigMod1}(e) and (f)). The results show that the sensitivity improves towards an optimum when the microwave and the optical pumping power decrease. Further lowering of these parameters then degrades the sensitivity because (i) the rate of detected photons decreases with the optical pumping power and (ii) the contrast quickly decreases with the microwave power. Using the values of $\Gamma_{p}^{\infty}$, $\Gamma_{c}^{\infty}$, $\Theta$ and $\mathcal{R}^{\infty}$ previously obtained, we can finally compute a two dimensional plot of the magnetic field sensitivity, as shown in Fig.~\ref{FigMod1}(g). An optimal sensitivity $\eta_{B}\approx 2 \ \mu$T/$\sqrt{\rm Hz}$ is obtained, which corresponds to the best sensitivity that can be achieved by using continuous optically detected ESR spectroscopy of single NV defects in a high purity CVD-grown diamond crystal.

\section{Magnetic field sensitivity using pulsed-ESR spectroscopy}

\begin{figure}[t]
\includegraphics[width = 9cm]{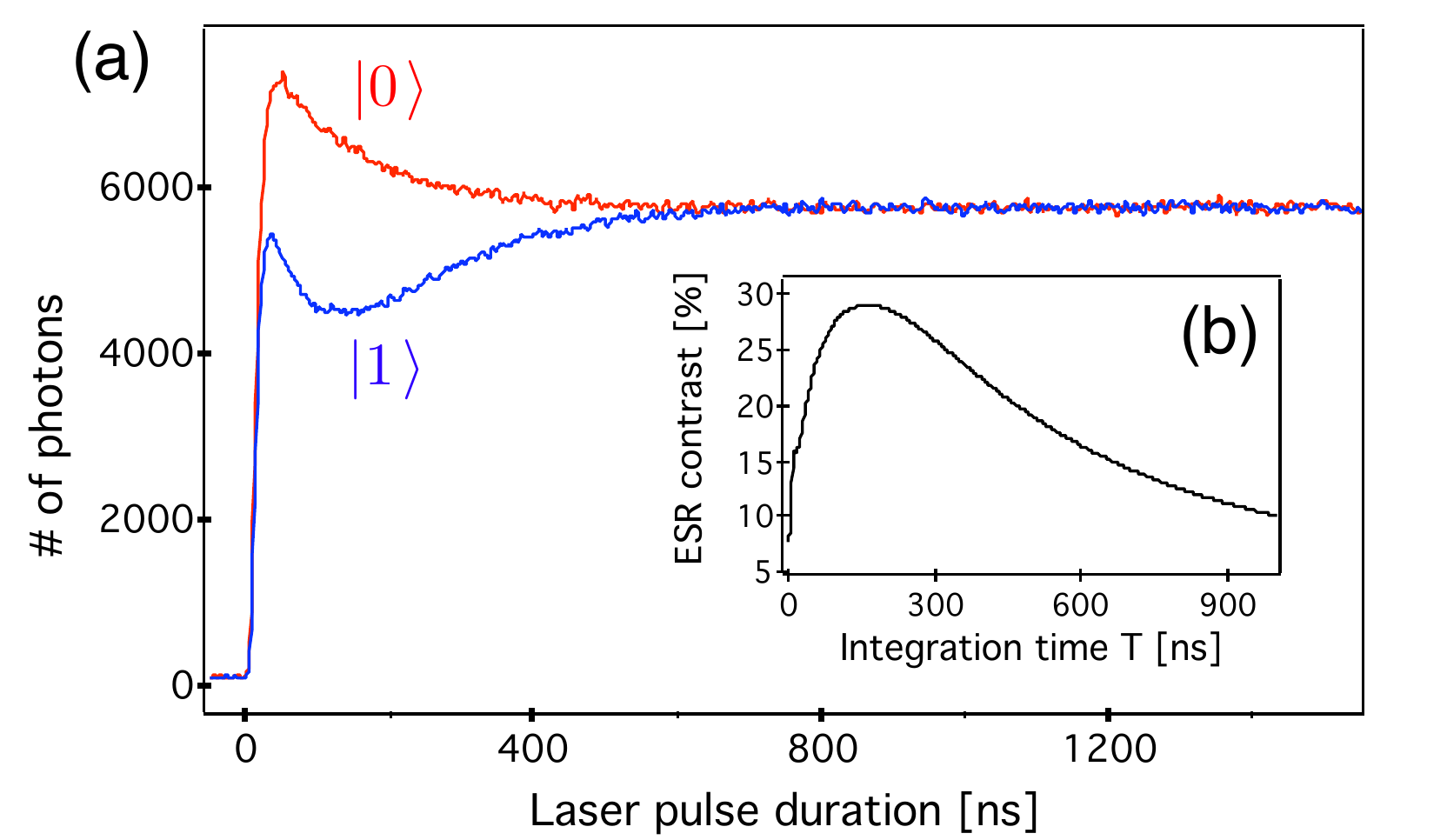}
 \caption{(color online). (a)-Time-resolved PL response of a single NV defect initially prepared either in the $\left|0\right.\rangle$ state (upper trace in red) or in the $\left|1\right.\rangle$ state (lower trace in blue). Preparation in $\left|0\right.\rangle$ is done by optical pumping with a $2 \ \mu$s laser pulse. A subsequent resonant MW $\pi$-pulse (50 ns) is applied for initialization in the $\left|1\right.\rangle$ state. The time binning is $1$ ns. (b)-ESR contrast $\mathcal{C}(T)$ as a function of the integration time $T$.}
 \label{Fig2}
 \end{figure}

We now demonstrate a simple method allowing to fully eliminate power broadening of the ESR linewidth while preserving a high contrast, thus enhancing the magnetic field sensitivity.\\
\indent For that purpose, we first analyse the time-resolved~PL during a read-out laser pulse for a single NV defect initially prepared either in state $\left|0\right.\rangle$ by optical pumping, or in state $\left|1\right.\rangle$ by applying an additional resonant microwave $\pi$-pulse (Fig.~\ref{Fig2}(a)). If the initial state is $\left|0\right.\rangle$, a high PL signal is initially observed which decays to a steady-state value for which some populations are trapped in the metastable state owing to residual ISC to the metastable state from the $m_{s}=0$ excited state (Fig.~\ref{Fig1}(a)). We note that such processes were not taken into account in the model discussed in the previous section. In order to predict precisely the spin dependence of time-resolved PL, a five-level model of the NV defect has to be developed, as described in Refs.~\cite{Manson_PRB2006,Hanson_NJP2011}. If the initial state is $\left|1\right.\rangle$, the time-resolved PL signal rapidly decays to a low level owing to fast ISC to the metastable state (Fig.~\ref{Fig1}(a)). Since the metastable state preferentially decays to the $\left|0\right.\rangle$ state, the low PL level then decays towards the steady-state value within the metastable state lifetime. 

\indent We note $\mathcal{N}_{0}(T)$ (resp. $\mathcal{N}_{1}(T)$) the total number of collected photons during an integration time $T$ for a single NV defect initially prepared in state $\left|0\right.\rangle$ (resp. $\left|1\right.\rangle$). The effective signal used to discriminate between the different spin sublevels is given by $\mathcal{S}(T)=\mathcal{N}_{0}(T)-\mathcal{N}_{1}(T)$. In particular, the ESR contrast $\mathcal{C}(T)$ is defined by :
\begin{equation}
\mathcal{C}(T)=\frac{\mathcal{N}_{0}(T)-\mathcal{N}_{1}(T)}{\mathcal{N}_{0}(T)} \ .
\label{contraste}
\end{equation}
Since the signal $\mathcal{S}(T)$ saturates when spin populations reach their steady-state values, the ESR contrast decreases for an integration time longer than the metastable state lifetime (Fig.~\ref{Fig2}(b)). \\
 \begin{figure}[t]
 \includegraphics[width = 9cm]{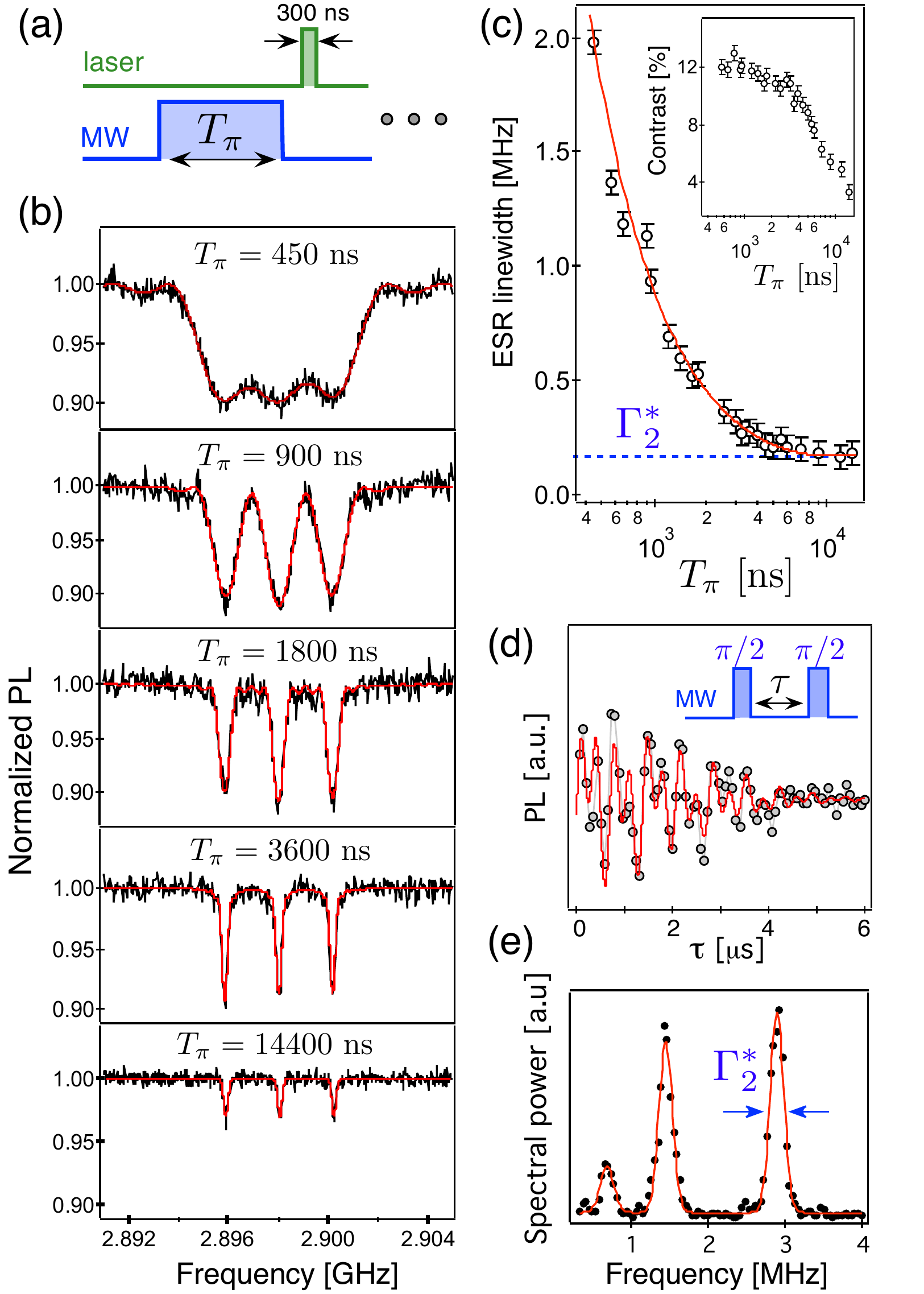}
 \caption{(color online). (a)-Pulse sequence used to eliminate power broadening effects. (b)-Pulsed-ESR spectra recorded for different values of the $\pi$-pulse duration $T_{\pi}$. The three resonances correspond to the three hyperfine components associated with the $^{14}$N nuclear spin. The laser power corresponds to $s=1.2$. (c)-ESR linewidth as a function of the $\pi$-pulse duration plotted in log scale. The solid line is the convolution between a sinc function of width $\Gamma_{\pi}\propto T_{\pi}^{-1}$ and a Gaussian function of width $\Gamma_{2}^{*}=2 \times 10^{5}$~s$^{-1}$. The inset shows the evolution of the ESR contrast as a function of the $\pi$-pulse duration $T_{\pi}$. (d)-Ramsey fringes recorded for the same NV defect with a microwave detuning of $0.7$ MHz from the central hyperfine line of the ESR spectrum. The red solid line is data fitting with the function $\exp[(\tau/T_{2}^{*})^{2}]\sum_{i=1}^{3}\cos(2\pi f_{i}\tau)$, where $f_{i}$ values are the microwave detunings from each hyperfine component of the spectrum. A value $T_{2}^{*}=3.0\pm 0.2 \ \mu$s is achieved. (e)-Fourier-transform spectrum of Ramsey fringes. Solid lines are data fitting with Gaussian functions, leading to $\Gamma_{2}^{*}=\left(2.08\pm 0.05\right) \times 10^{5}$~s$^{-1}$. }
 \label{Fig3}
 \end{figure}
\indent According to this result, power broadening can be fully eliminated in ESR spectra by performing pulsed ESR in dark condition with the simple sequence depicted in Fig.~\ref{Fig3}(a). A microwave $\pi$-pulse is followed by a laser pulse used both for spin-state read-out with a high contrast and to achieve an efficient preparation of the NV defect in the $\left|0\right.\rangle$ state for the next microwave $\pi$-pulse. The duration of the read-out laser pulse was set to $T_{L}=300$~ns and each laser pulse was followed by a $1 \ \mu$s waiting time in order to ensure the relaxation of steady state populations trapped in the metastable state towards the ground state $\left|0\right.\rangle$ before applying the next microwave $\pi$-pulse. ESR spectra were then recorded by continuously repeating the sequence while sweeping the $\pi$-pulse frequency and recording the PL intensity (Fig.~\ref{Fig3}(b)). Since spin rotations are induced in dark condition, power broadening from the laser is fully cancelled and the optical power can be set above the NV defect saturation ($s>1$).\\
\indent In this experiment, the ESR linewidth is given by the Fourier transform of the product of the $\pi$-pulse rectangular-shaped profile of duration $T_{\pi}$ by the inhomogeneous Gaussian profile of the NV defect electron spin, characterized by its coherence time $T_{2}^{*}$. This corresponds to the convolution of a sinc function (width $\Gamma_{\pi}\propto T_{\pi}^{-1}$) with a Gaussian function (width $\Gamma_{2}^{*}\propto T_{2}^{*-1}$). If $\Gamma_{\pi}\gg \Gamma_{2}^{*}$, each resonance of the ESR spectrum can be fitted by sinc functions, with a power-broadened linewidth $\Gamma \propto T_{\pi}^{-1}$ (Fig.~\ref{Fig3}(b)). By increasing the $\pi$-pulse duration, the linewidth becomes sharper and reaches the inhomogeneous linewidth $\Gamma_{2}^{*}\approx 2\times 10^{5}$~s$^{-1}$ when $T_{\pi}\approx T_{2}^{*}$. In this situation power broadening has been fully cancelled in the experiment and the data can be well fitted by a Gaussian profile. On the other hand, we note that the ESR contrast is not significantly altered until $T_{\pi}\approx T_{2}^{*}$. However, if $T_{\pi}$ is further increased, the linewidth remains limited by $\Gamma_{2}^{*}$ while the contrast begins to decrease (see inset in Fig.~\ref{Fig3}(c)).\\
\indent In order to verify that the inhomogeneous linewidth $\Gamma_{2}^{*}$ is indeed achieved in pulsed-ESR spectroscopy, Ramsey fringes were recorded by using the usual sequence consisting in two microwave $\pi/2$-pulses separated by a variable free evolution duration $\tau$ (Fig.~\ref{Fig3}(d))~\cite{Gopi_NatMat2009}. Data fitting of the free induction decay (FID) signal leads to a coherence time $T_{2}^{*}=3.0\pm 0.2 \ \mu$s of the NV defect electron spin and its Fourier transform spectrum exhibits a Gaussian profile with a linewidth $\Gamma_{2}^{*}=\left(2.08\pm 0.05\right) \times 10^{5}$~s$^{-1}$, as measured using pulsed ESR spectroscopy. \\

\begin{figure}[t]
 \includegraphics[width = 8.7cm]{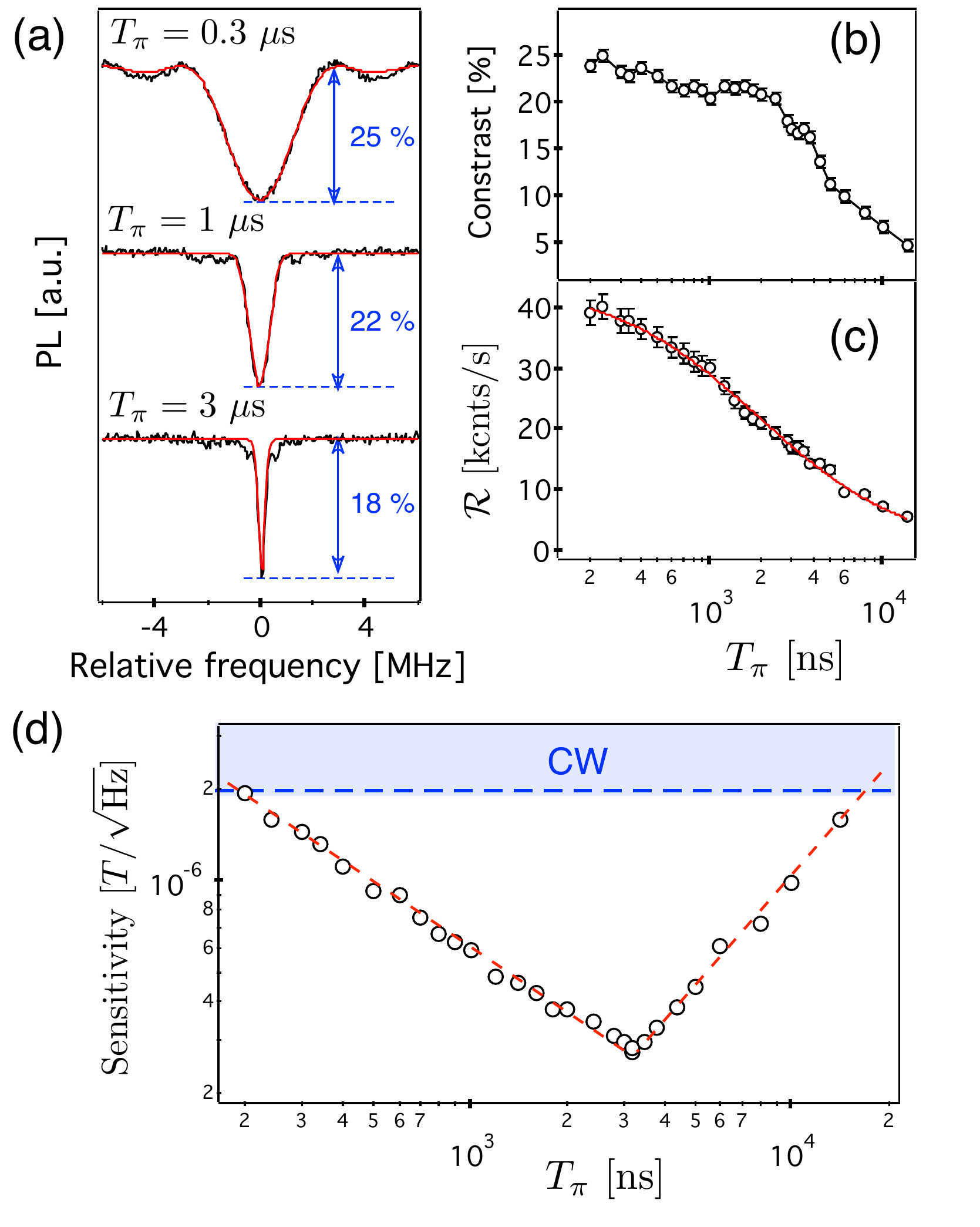}
 \caption{(color online). (a)-Pulsed-ESR spectra recorded at the excited-state LAC for different values of the $\pi$-pulse duration $T_{\pi}$. (b)-ESR contrast as a function of $T_{\pi}$. (c)-Averaged rate of detected photons $\mathcal{R}$ measured while running the pulsed-ESR sequence as a function of $T_{\pi}$. The solid line is data fitting with the function $\mathcal{R}=\mathcal{R}_{0}\times\frac{T_{L}}{T_{S}}$ where $\mathcal{R}_{0}$ is the rate of detected photons for a continuous laser excitation and $T_{S}$ the total duration of the pulse sequence, including initialization, $\pi$-pulse rotation and spin state read-out. (d)-Corresponding magnetic field sensitivity plotted in log-log scale using eq.~(\ref{sensib}). An enhancement by roughly one order-of-magnitude is achieved compared to continuous (CW) ESR spectroscopy. }
 \label{Fig6}
 \end{figure}

\indent We now compare the magnetic field sensitivity of pulsed and continuous ESR spectroscopy. For that purpose, all the measurements were reproduced at the excited-state LAC (Fig.~\ref{Fig6}). From a set of data including the ESR linewidth, the contrast and the averaged rate of detected photons $\mathcal{R}$ measured while running the pulsed-ESR sequence (Fig.~\ref{Fig6}(a) to (c)), the shot-noise-limited magnetic field sensitivity $\eta_{B}$ was estimated as a function of the $\pi$-pulse duration using eq.~(\ref{sensib}). As shown in Fig.~\ref{Fig6}(d), the sensitivity improves until an optimum $\eta_{B}\approx 300$~nT$/\sqrt{\rm Hz}$ when $T_{\pi}\approx T_{2}^{*}$. Further increase of $T_{\pi}$ impairs the sensitivity since the ESR contrast decreases significantly. Even if the rate of defected photons is limited by the duty cycle of the laser pulses in the ESR sequence, the magnetic field sensitivity is improved by roughly one order of magnitude in comparison to continuous ESR spectroscopy (Fig.~\ref{Fig6}(d)). We note that this sensitivity could be further enhanced through the conditional manipulation of the nitrogen nuclear spin of the NV defect at the excited-state LAC~\cite{Steiner_PRB2010}.\\
\indent For $T_{\pi}\approx T_{2}^{*}$, the rate of detected photons can be approximated by $\mathcal{R}\approx\mathcal{R}_{0}T_{L}/T_{2}^{*}$, where $\mathcal{R}_{0}$ is the rate of detected photons for a continuous laser excitation. Since $\Delta\nu=\Gamma_{2}^{\star}=\frac{2\sqrt{\ln 2}}{\pi T_{2}^{*}}$, the magnetic field sensitivity can then be written as 
\begin{equation}
\eta_{B}\approx \sqrt{2e}\times \frac{\hbar}{g\mu_{B}}\times\frac{1}{\mathcal{C}\sqrt{\mathcal{R}_{0}T_{L}}}\times\frac{1}{\sqrt{T_{2}^{*}}} \ .
\end{equation}
Within a numerical factor, this formula is similar to the one obtained for the optimum sensitivity of a magnetometer based on a single NV defect while using a Ramsey-type sequence~\cite{Taylor_NatPhys2008}. Furthermore, we note that the ESR contrast is reduced by approximately $20\%$ for $T_{\pi}\approx T_{2}^{*}$ (Fig.~\ref{Fig6}(b)). Such an effect also slightly degrades the magnetic field sensitivity compared to a Ramsey type experiment.

\section{Conclusion}
We have reported a systematic study of the magnetic field sensitivity of a magnetic sensor consisting of a single NV defect in diamond, by using optically detected ESR spectroscopy in continuous and pulsed regimes. By using a simple pulsed ESR sequence based on the repetitive excitation of the NV defect with a resonant microwave $\pi$-pulse, we have shown that power broadening of the ESR linewidth can be fully suppressed, leading to an enhancement of the magnetic field sensitivity by roughly one order of magnitude in comparison to continuous ESR spectroscopy. Apart from magnetometry applications, the reported pulsed ESR scheme appears as a useful tool for the study of weak hyperfine interactions of the NV defect with nearby nuclear spins in the diamond matrix~\cite{Smeltzer_NJP2011}.

\section*{Appendix}
\indent Using the notations introduced in the main text and eq.~(\ref{Liouville}), Bloch equations of the {\it closed} two-level system shown in Fig.~\ref{Fig1} read as :
\begin{equation}
\frac{d\sigma_{11}}{dt}=i\frac{\Omega_{R}}{2} \left[\sigma_{10}-\sigma_{01}\right]-\Gamma_{1}\left[ \sigma_{11}-\sigma_{00} \right] - \Gamma_{p} \sigma_{11} 
\end{equation}
\begin{equation}
\frac{d\sigma_{01}}{dt}=i\left[\omega_{0}-\omega_{m}\right]\sigma_{01}-i\frac{\Omega_{R}}{2}\left[\sigma_{11}-\sigma_{00}\right]-\Gamma_{2}\sigma_{01} \ ,
\end{equation}
where $\Gamma_{2}=\Gamma_{2}^{\star}+\Gamma_{c}$. The steady state solutions of the populations $\sigma_{ii}^{st}$ are then given by 
\begin{eqnarray}
\sigma_{11}^{st}&=&\frac{\displaystyle\frac{\Gamma_{1}}{2\Gamma_{1}+\Gamma_{p}}\left( [\omega_{0}-\omega_{m}]^{2}+\Gamma_{2}^{2}\right)+\displaystyle\frac{\Gamma_{2}\Omega_{R}^{2}/2}{2\Gamma_{1}+\Gamma_{p}}}{[\omega_{0}-\omega_{m}]^{2}+\Gamma_{2}^{2}+\displaystyle\frac{\Gamma_{2}\Omega_{R}^{2}}{2\Gamma_{1}+\Gamma_{p}}}\\
\sigma_{00}^{st}&=&1-\sigma_{11}^{st} \ .
\label{steady}
\end{eqnarray}
\indent When the optical pumping is switched off ($\Gamma_{p}=0$), $\sigma_{11}^{st}=\sigma_{00}^{st}=1/2$. On the other hand if the microwave power is off ($\Omega_{R}=0$), we obtain $\sigma_{11}^{st}=\Gamma_{1}/(2\Gamma_{1}+\Gamma_{p})$. Consequently, if $\Gamma_{p}\gg\Gamma_{1}$ then  $\sigma_{11}^{st}\approx0$, corresponding to the well known optically-induced polarization of the NV defect in state $\left|0\right.\rangle$. Conversely, if the optical pumping is such that $\Gamma_{p}\ll\Gamma_{1}$, then $\sigma_{11}^{st}=\sigma_{00}^{st}=1/2$. Intermediate cases could be investigated by studying large ensembles of NV defects.\\
\indent Using eq.~(\ref{PL}) and~(\ref{steady}), the contrast reads as
\begin{equation}
\mathcal{C}=\frac{1}{2}\times \frac{(\alpha-\beta)\Gamma_{p}}{(\alpha+\beta)\Gamma_{1}+\alpha\Gamma_{p}} \times \frac{\Omega_{R}^{2}}{\Omega_{R}^{2}+\Gamma_{2}(2\Gamma_{1}+\Gamma_{p})} \ .
\label{ContrastCW}
\end{equation}

As expected, the contrast vanishes if (i) $\alpha=\beta$, (ii) when $\Omega_{R}\rightarrow 0$ and (iii) $\Gamma_{p}\rightarrow 0$.\\ 
The associated ESR linewidth $\Delta\nu$ is given by
\begin{equation}
\label{Bloch}
\Delta\nu=\frac{1}{2\pi}\sqrt{\Gamma^{2}_{2}+\frac{\Omega_{R}^{2}\Gamma_{2}}{2\Gamma_{1}+\Gamma_{p}}} \ .
\end{equation}
By considering $s>10^{-2}$, corresponding to $\Gamma_{p}\gg \Gamma_{1}$ and $\Gamma_{2}= \Gamma_{c}$, the ESR contrast and the linewidth are finally given by eq.~(\ref{Cfinal}) and~(\ref{Linewidth}) of the main text. The shot-noise limited magnetic field sensitivity can then be inferred by using  eq.~(\ref{sensib}). We note that for a fixed value of the saturation parameter $s$, the Rabi frequency which optimizes the magnetic field sensitivity is given by 
\begin{equation}
\label{opt}
\Omega_{R}=\sqrt{2\Gamma_{p}^{\infty}\Gamma_{c}^{\infty}}\times \frac{s}{1+s} \ .
\end{equation}

\begin{acknowledgments}
The authors acknowledge F.~Grosshans, P.~Neumann and F.~Jelezko for fruitful discussions. This work was supported  by the Agence Nationale de la Recherche (ANR) through the project D{\sc iamag}, by C'Nano \^Ile-de-France and by RTRA-Triangle de la Physique (contract 2008-057T).
\end{acknowledgments}

%\newpage

\end{document}